\documentclass[prd,twocolumn,superscriptaddress,floatfix,nopacs,preprintnumbers,nofootinbib]{revtex4-2}
\usepackage[utf8]{inputenc}

\usepackage[caption = false]{subfig}
\usepackage[normalem]{ulem}
\usepackage{xcolor}

\usepackage{mathrsfs}
\usepackage{physics}
\usepackage{amssymb,bm}
\usepackage{amsmath}
\usepackage{mathtools}
\usepackage{slashed}
\usepackage{graphics}
\usepackage{graphicx}
\usepackage{tikz}

\newcommand{\nc}{N_\mathrm{c}}

\newcommand{\jpsi}{$\mathrm{J}/\psi$ }
\newcommand{\bt}{\mathbf{b}}
\newcommand{\rt}{\mathbf{r}}

\newcommand{\xpom}{x_\mathbb{P}}

\newcommand{\lqcd}{\Lambda_\mathrm{QCD}}

\newcommand{\Deltat}{\mathbf{\Delta}}

\usepackage[breaklinks,colorlinks,citecolor=citcolor,urlcolor=blue,linkcolor=lcolor]{hyperref}

\definecolor{lcolor}{rgb}{0.5,0,0}
\definecolor{citcolor}{rgb}{0,0.3,0.0}

\begin{document}

\author{Heikki M\"{a}ntysaari}
\email{heikki.mantysaari@jyu.fi}
\affiliation{
Department of Physics, University of Jyväskylä,  P.O. Box 35, 40014 University of Jyväskylä, Finland
}
\affiliation{
Helsinki Institute of Physics, P.O. Box 64, 00014 University of Helsinki, Finland
}

\author{Farid Salazar}
\affiliation{Institute for Nuclear Theory, University of Washington, Seattle WA 98195-1550, USA}
\affiliation{Nuclear Science Division, Lawrence Berkeley National Laboratory, Berkeley, California 94720, USA}
\affiliation{Physics Department, University of California, Berkeley, California 94720, USA}
\affiliation{Department of Physics and Astronomy, University of California, Los Angeles, California 90095, USA}
\affiliation{Mani L. Bhaumik Institute for Theoretical Physics, University of California, Los Angeles, California 90095, USA}

\author{Bj\"orn Schenke}
\affiliation{Physics Department, Brookhaven National Laboratory, Upton, NY 11973, USA}

\title{Energy dependent nuclear suppression from gluon saturation\\ in exclusive vector meson production}

\begin{abstract}
    We calculate the exclusive \jpsi photoproduction cross section at high energies from the Color Glass Condensate approach. 
    The results are compared to the center-of-mass energy dependent $\gamma+A\to\mathrm{J}/\psi+A$ cross sections extracted from measurements in ultra peripheral heavy ion collisions at RHIC and LHC. We predict strong saturation-driven nuclear suppression at high energies, while LHC data prefers even stronger suppression. We explore the effect of nucleon shape fluctuations on the nuclear suppression in the coherent and incoherent cross sections, and show that the most recent measurement of the $|t|$-differential incoherent \jpsi cross section prefers large event-by-event fluctuations of the nucleon substructure in heavy nuclei, comparable to that found for a free proton. 
\end{abstract}

\maketitle 

\section{Introduction}

Exclusive vector meson production in high-energy photon-nucleus collisions is an especially powerful tool to probe the nuclear wave function at small longitudinal momentum fraction. This is because in an exclusive process at least two gluons need to be exchanged, rendering the process very sensitive to the target structure~\cite{Brodsky:1994kf,Guzey:2020ntc,Eskola:2022vaf}. Additionally, the possibility to measure the total transverse momentum transfer provides access to the (event-by-event fluctuating) spatial distribution of nuclear matter in the target nucleus at small momentum fraction $\xpom$~\cite{Klein:2019qfb,Mantysaari:2020axf}. Finally, the probe is a photon whose structure can be understood perturbatively, and the kinematics in the process can be determined completely. Consequently, vector meson production processes will play a central role at the future EIC~\cite{AbdulKhalek:2021gbh} and LHeC/FCC-he~\cite{LHeC:2020van} nuclear deep inelastic scattering (DIS) facilities when looking for signals of non-linear saturation effects.

Saturation effects are expected to be encountered in heavy nuclei at high energies where the parton densities become so large that gluon emission and gluon recombination processes balance each other. At such high densities (or energies) it is convenient to describe QCD dynamics using the Color Glass Condensate (CGC)~\cite{Gelis:2010nm,Kovchegov:2012mbw} effective theory. CGC calculations have been extensively applied to many different collider experiments (for a review, see e.g.~Ref.~\cite{Morreale:2021pnn}). However, at the moment there are no unambiguous signals of gluon saturation observed. To change this situation, it is important to focus on the study of clean processes that are especially sensitive to saturation effects, such as exclusive vector meson production at the highest achievable energies~\cite{Mantysaari:2017slo}. Here, the \jpsi production process is intriguing, as the mass of the \jpsi is large enough to ensure perturbative stability, but low enough to keep the process sensitive to the non-linear dynamics that is important for momenta lower than the saturation scale.

Before the future $e+A$ colliders are realized, it is possible to study very high-energy photoproduction processes in ultra peripheral collisions (UPCs)~\cite{Bertulani:2005ru,Klein:2019qfb} at RHIC and the LHC, see e.g.~Refs.~\cite{ALICE:2021gpt,CMS:2023snh,LHCb:2022ahs} for some recent \jpsi production cross section measurements. Although in principle one can access very small $\xpom\sim 10^{-5}$ at the LHC at forward rapidities, in ultra peripheral $\mathrm{Pb}+\mathrm{Pb}\to\mathrm{J}/\psi+\mathrm{Pb}+\mathrm{Pb}$ there is generally a two-fold ambiguity in the kinematics: \jpsi production at a given rapidity could result from a high-energy photon emitted from the first nucleus scattering off a small-$\xpom$ gluon from the other nucleus, or a low energy photon from the second nucleus scattering off a large $\xpom$ gluon from the first nucleus. Because the high-energy photon flux is heavily suppressed, the sensitivity to the very small-$\xpom$ structure is limited. 

Recently ALICE~\cite{ALICE:2023jgu}, CMS~\cite{CMS:2023snh} and STAR~\cite{STAR:2023gpk,STAR:2023nos} collaborations have extracted the center-of-mass-energy dependence of the $\gamma+A\to\mathrm{J}/\psi+A$ cross section from the measured \jpsi production cross section in ultra peripheral collisions using the method proposed in Ref.~\cite{Guzey:2013jaa}. Thanks to these developments, it has become possible to study photon-nucleus scattering at energies up to $W\sim 1$~TeV, where the nucleus is probed at $\xpom\sim 10^{-5}$. These developments enable clean studies of gluon saturation phenomena in a unique kinematical domain where one can expect saturation effects in heavy nuclei to be strong.

\jpsi production in ultra peripheral collisions (UPCs) has been extensively studied within the CGC framework, see e.g.~Refs.~\cite{Lappi:2013am,Mantysaari:2022sux,Goncalves:2017wgg,Bendova:2020hbb}.
The purpose of this paper is to extend the UPC results presented in Ref.~\cite{Mantysaari:2022sux} to the high-energy photon-nucleus collisions covered by the recent photoproduction measurements. We present the state-of-the-art CGC predictions for the energy dependence of the \jpsi photoproduction cross sections and nuclear suppression factors to determine the compatibility of the applied gluon saturation picture with the new data, providing access to very small-$\xpom$ kinematics in a process which is both experimentally and theoretically clean. 
Furthermore, we also present a comparison to the new UPC measurement of the $t$-dependent incoherent cross section that has become available since the publication of Ref.~\cite{Mantysaari:2022sux}. 
This paper is organized as follows: The applied CGC framework is briefly reviewed in Sec.~\ref{sec:setup}, and comparisons to new RHIC and LHC \jpsi photoproduction data are shown in Sec.~\ref{sec:results}. Conclusions are presented in Sec.~\ref{sec:conclusions}.

\section{Exclusive vector meson production in the Color Glass Condensate}
\label{sec:setup}

We calculate exclusive vector meson production using the same setup as in Ref.~\cite{Mantysaari:2022sux}, which we briefly summarize here. At high energies, the process factorizes such that first the virtual photon fluctuates into a quark-antiquark dipole at leading order (see Ref.~\cite{Mantysaari:2021ryb,Mantysaari:2022kdm} for an extension to NLO), and then the quarks propagate eikonally through the target color field before forming a vector meson. As such, the coherent cross section for $\gamma+A\rightarrow \mathrm{J}/\psi+A$ can be written as~\cite{Mantysaari:2020axf,Good:1960ba,Miettinen:1978jb,Caldwell:2010zza} 
\begin{equation}
    \frac{\dd \sigma}{\dd t} = \frac{1}{4\pi} \left|\langle \mathcal{A}\rangle_{\xpom} \right|^2,
\end{equation}
and the incoherent cross section reads
\begin{equation}
\label{eq:incohxs}
    \frac{\dd \sigma^{\gamma A}}{\dd t} = \frac{1}{4\pi} \left[ \langle|\mathcal{A}|^2\rangle_{\xpom} - |\langle \mathcal{A}\rangle_{\xpom}|^2 \right].
\end{equation}
Here $\langle\rangle_{\xpom}$ refers to the average over target color field configurations at the given $\xpom$, and the scattering amplitude $\mathcal{A}$ is
\begin{align}
     -i{\mathcal{A}} = \int \dd[2]{\rt} \dd[2]\bt  \int_0^1 \frac{\dd{z}}{4\pi}  &[\Psi_V^* \Psi_\gamma](Q^2,\rt,z) \nonumber \\
     &\times e^{-i\bt\cdot\Deltat}  N(\rt,\bt,z) \,.
\end{align}
In this work, we only consider photoproduction processes where $Q^2=0$.

All information about the target structure is encoded in the two-point Wilson line operator
\begin{align}
    &N(\rt,\bt,z)  \nonumber \\
    & = 1 - \frac{1}{\nc} \tr \left[ V\left(\bt + (1-z)\rt\right) V^\dagger\left(\bt - z\rt\right) \right]. 
\end{align}
Here $\rt$ and $\bt$ are the dipole size (and orientation) and impact parameter (center of the $q\bar q$ dipole), and the dependence on the longitudinal momentum fraction $z$  takes into account the non-forward phase (or the fact that the Fourier conjugate to the momentum transfer $\Deltat$ is the center-of-mass of the dipole)~\cite{Hatta:2017cte,Mantysaari:2020lhf}. Explicit expressions for the photon and vector meson wave functions $\Psi_\gamma$ and $\Psi_V$ can be found from Ref.~\cite{Kowalski:2006hc}. For the \jpsi we use the Boosted Gaussian model with parameters constrained in Ref.~\cite{Mantysaari:2018nng}. The \jpsi wave function is not accurately known~\cite{Lappi:2020ufv}, but this uncertainty mostly affects the overall normalization of the \jpsi production cross section and is to a large extent removed when the free parameters are constrained by the $\gamma+p\to\mathrm{J}/\psi+p'$ photoproduction data we discuss next.

The Wilson lines at the initial $\xpom=0.01$ are obtained from the McLerran-Venugopalan model~\cite{McLerran:1993ni,McLerran:1993ka}. The energy ($\xpom$) dependence is obtained by solving the JIMWLK evolution equations ~\cite{Mueller:2001uk}. The free parameters describing the proton saturation scale, the scale at which the coordinate space running coupling is evaluated, and the event-by-event fluctuating proton geometry are determined by fitting the \jpsi photoproduction cross section in $\gamma+p$ collisions as measured by H1~\cite{H1:2005dtp,H1:2013okq}, ZEUS~\cite{ZEUS:2002wfj}, ALICE~\cite{ALICE:2014eof,ALICE:2018oyo} and LHCb~\cite{LHCb:2014acg, LHCb:2018rcm} (see also e.g.~Refs.~\cite{Mantysaari:2022ffw,Mantysaari:2016jaz,Mantysaari:2016ykx,Mantysaari:2018zdd,Kumar:2021zbn,Cepila:2016uku} for further extractions of the event-by-event fluctuating geometry). These parameters are determined separately for the case where the proton has no substructure but only color charge fluctuations (referred to as ``CGC'' in this work), and for the case where the proton consists of three fluctuating hot spots (``CGC+shape fluct.''). We note that the coherent cross section for $\gamma+p\to\mathrm{J}/\psi+p$ is practically identical in both setups at all center-of-mass energies probed at the LHC as shown in Ref.~\cite{Mantysaari:2022sux}. 

When calculating cross sections for ultra peripheral collisions ($\mathrm{Pb}+\mathrm{Pb}\to \mathrm{J}/\psi+\mathrm{Pb}+\mathrm{Pb}$) the photon-nucleus cross section is multiplied by an equivalent photon flux integrated over nucleus-nucleus impact parameter $|\mathbf{B}|>2R_A$ where we use $R_A=6.62$ fm for Pb as in Ref.~\cite{Mantysaari:2022sux}. The UPC observables integrated over momentum transfer $t$ considered in this work are not sensitive to the interference effect (present because both nuclei can act as photon sources) or to the non-zero but small photon transverse momentum, so these effects are not included here.

\section{Results}
\label{sec:results}
\subsection{Vector meson photoproduction in UPCs}
\label{sec:default_upc_comparison}

\begin{figure}
           \includegraphics[width=\columnwidth]{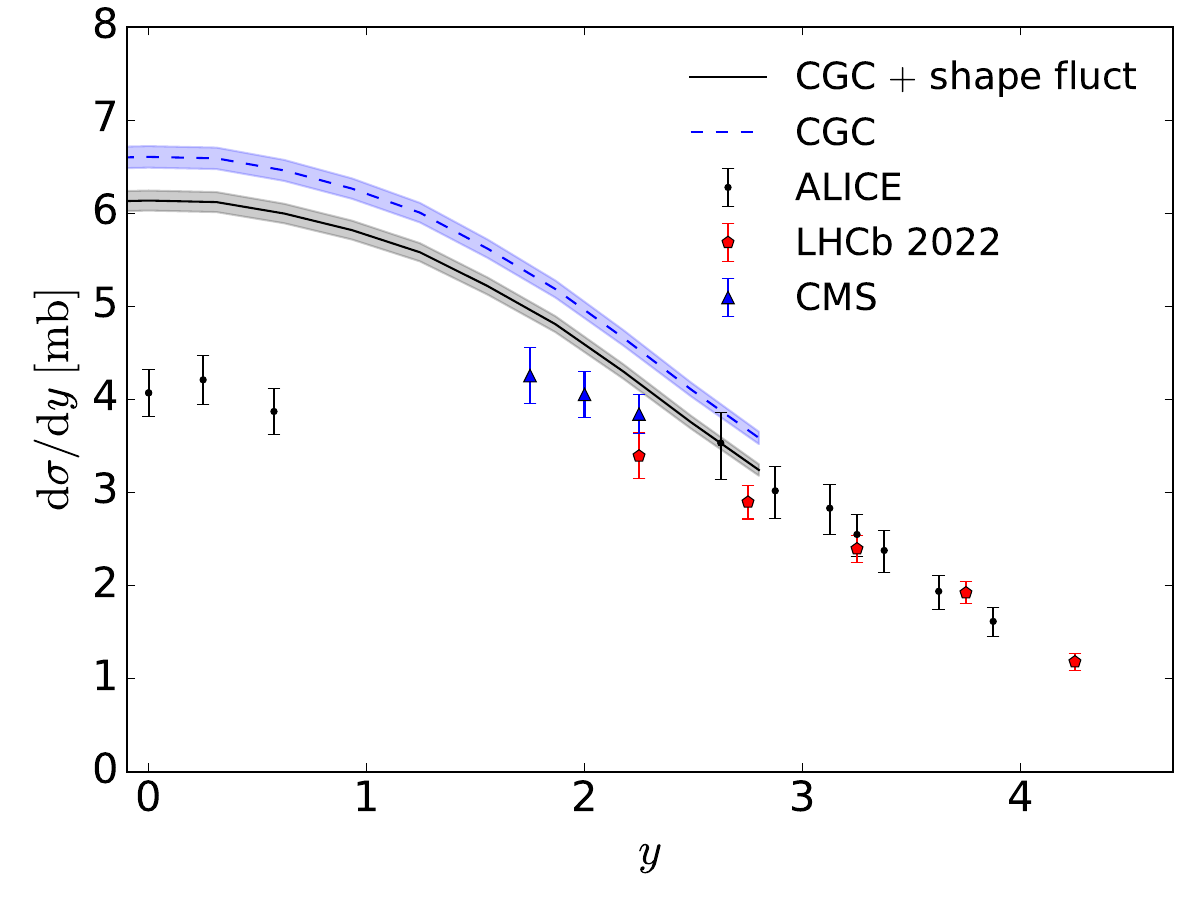}
           \begin{tikzpicture}[overlay,remember picture]
     \node at (-0.7cm,1.5cm){Coherent $\mathrm{Pb}+\mathrm{Pb}\to\mathrm{J}/\psi+\mathrm{Pb}+\mathrm{Pb}$};
     \end{tikzpicture}
           \caption{Coherent \jpsi photoproduction cross section in ultra peripheral $\mathrm{Pb}+\mathrm{Pb}$ collisions compared to the ALICE~\cite{ALICE:2021gpt}, CMS~\cite{CMS:2023snh}, and LHCb~\cite{LHCb:2022ahs} data. The bands represent the statistical uncertainty of the calculation.}
           \label{fig:coh_5020_nolhcb2021}
    \end{figure}
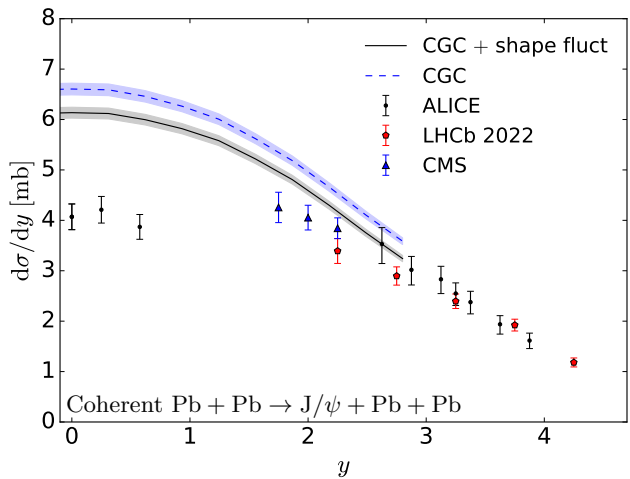

Before discussing photon-nucleus cross sections, we first compute coherent \jpsi production in ultra peripheral Pb+Pb collision where there is no additional uncertainty related to the extraction of the photonuclear cross section. On the other hand, sensitivity to the very small $\xpom$ structure of the nucleus is much more limited.

The coherent \jpsi production cross section as a function of \jpsi rapidity is shown in Fig.~\ref{fig:coh_5020_nolhcb2021}. As discussed above, we use the same setup as in Ref.~\cite{Mantysaari:2022sux} summarized in Sec.~\ref{sec:setup}, and as such the CGC predictions obtained with and without nucleon shape fluctuations are identical to those presented in our previous publication. Here we compare to newly available data from the CMS Collaboration~\cite{CMS:2023snh}, covering a previously unexplored rapidity range, in addition to the most recent data from LHCb~\cite{LHCb:2022ahs} and ALICE~\cite{ALICE:2021gpt}. 

The inclusion of the new CMS dataset does not significantly modify the conclusions already presented in Ref.~\cite{Mantysaari:2022sux}. The cross section is slightly smaller when the proton shape fluctuations are included. This is because the non-linear effects are stronger in the fluctuating case where there are regions with larger local saturation scales, leading to more suppression. The rapidity dependence of the ALICE, LHCb, and CMS data is quite well reproduced by our calculation in the $|y|\gtrsim 1.5$ region, but the ALICE midrapidity data is significantly overestimated. 

Due to the two-fold ambiguity of the UPC kinematics, at $y\neq 0$ one probes the nucleus at two different values of $\xpom=\frac{M_V}{\sqrt{s}}e^{\pm y}$ where $M_V$ is the vector meson mass and $\sqrt{s}$ the nucleon-nucleon center-of-mass energy. The CGC calculations in Fig.~\ref{fig:coh_5020_nolhcb2021} are limited to the region where $\xpom<0.01$, as the initial condition for the JIMWLK evolution is parametrized at $\xpom=0.01$. We note that the larger-$\xpom$ contribution dominates in the large-$y$ region, which means that our calculations agree with the LHC data well in the domain where the dominant contribution comes from the process with a relatively low photon-nucleon center-of-mass energy $W^2=\sqrt{s}M_V e^{-y}\lesssim(60\,\mathrm{GeV})^2$. Smaller $\xpom$ dominates at the lowest $y$ values in this observable, implying that our calculation increasingly underestimates the nuclear suppression as $\xpom$ decreases.
Comparisons to the $t$-integrated incoherent cross section and the coherent cross section data at $\sqrt{s}=2.76$~TeV are shown in Ref.~\cite{Mantysaari:2022sux}. There are no updates for these datasets.

\subsection{Vector meson production in photon-nucleus collisions}
\label{sec:photonuclear}

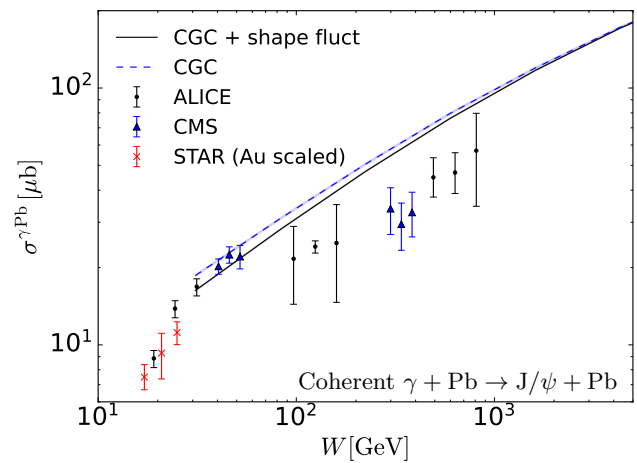
\begin{figure}
    \centering
    \includegraphics[width=\columnwidth]{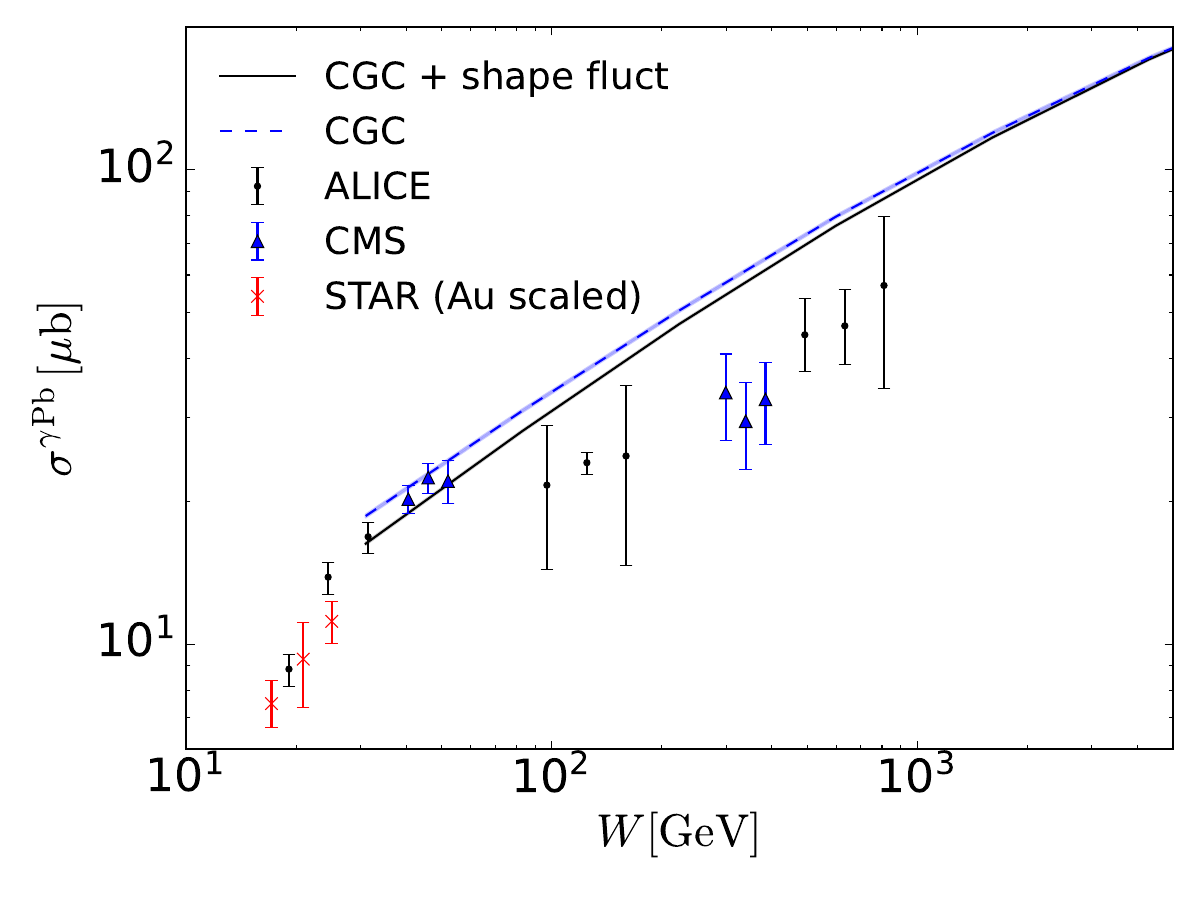}
    \begin{tikzpicture}[overlay,remember picture]
     \node at (1.8cm,1.7cm){Coherent $\gamma+\mathrm{Pb}\to\mathrm{J}/\psi+\mathrm{Pb}$};
     \end{tikzpicture}
        \caption{Center-of-mass energy dependence of the coherent \jpsi photoproduction cross section. The results are compared to the ALICE~\cite{ALICE:2023jgu}, CMS~\cite{CMS:2023snh} and scaled STAR~\cite{STAR:2023gpk,STAR:2023nos} data.}
        \label{fig:photonuclearjpsi}
\end{figure}

We now move to the main focus of this paper: Energy dependent diffractive vector meson production in photon-nucleus collisions. The two-fold ambiguity can be overcome and the photon-nucleus cross section extracted from the measured UPC cross section using the approach proposed in Ref.~\cite{Guzey:2013jaa}: UPCs with a different number of emitted forward neutrons correspond to different nucleus-nucleus distances and with that different photon fluxes. When the UPC cross section is measured in different neutron multiplicity classes, it becomes possible to solve for the $\gamma+A$ cross sections at different $\gamma$-nucleon center of mass energies $W$\footnote{Similarly one can compare photoproduction cross sections in peripheral and ultra peripheral processes as suggested in Ref.~\cite{Contreras:2016pkc}, but coherent production in a process with significant hadronic activity is theoretically more challenging to describe~\cite{Klein:2023zlf}.}. This procedure has been recently employed by the ALICE~\cite{ALICE:2023jgu}, CMS~\cite{CMS:2023snh} and STAR~\cite{STAR:2023gpk,STAR:2023nos} Collaborations to measure the photoproduction cross section for the $\gamma + \mathrm{Pb}\ (\mathrm{Au}) \to \mathrm{J}/\psi + \mathrm{Pb}\ (\mathrm{Au})$ scattering.

The coherent \jpsi photoproduction cross section as a function of the photon-nucleon center-of-mass energy $W$ is shown in Fig.~\ref{fig:photonuclearjpsi}. The results calculated for $\gamma+\mathrm{Pb}\to \mathrm{J}/\psi+\mathrm{Pb}$, again with and without nucleon shape fluctuations, are compared to the available ALICE, CMS and STAR data.
The datapoint at $W=125$ GeV with a very small uncertainty corresponds to midrapidity kinematics in UPCs at $\sqrt{s}=5020\,\mathrm{GeV}$ where there is no two-fold ambiguity and one can directly extract the $\gamma+A$ cross section.
The STAR measurements with gold targets are scaled to the photon-lead case by assuming that the $t$-integrated cross section scales as $A^{4/3}$~\cite{Mantysaari:2017slo}, which in this case corresponds to a multiplicative factor of $1.075$.

The measured $\gamma+\mathrm{Pb}$ cross section is well reproduced in the low center-of-mass energy $W\lesssim 100$ GeV region, but the high-energy cross sections are overestimated by up to 40\%. This is consistent with the result in Fig.~\ref{fig:coh_5020_nolhcb2021}: The UPC cross section at forward rapidities where the low-$W$ contribution dominates is well reproduced, but the midrapidity data corresponding to $W=125$~GeV is overestimated by $50\%$ (for the case with nucleon shape fluctuations). 
The slightly better agreement of our calculations with the photoproduction data compared to the midrapidity UPC measurement in Fig.~\ref{fig:coh_5020_nolhcb2021} is explained by the fact that the midrapidity data corresponds to any number of forward neutrons, but the $\gamma+\mathrm{Pb}$ data shown in Fig.~\ref{fig:photonuclearjpsi} is extracted from an independent measurement where different forward neutron classes are measured separately. Furthermore, there is also a few-percent difference in the photon flux used in our UPC setup~\cite{Mantysaari:2022sux} compared to that used by the ALICE collaboration.

Although the overall normalization of the cross section is overestimated in the high-energy region, our calculations capture well the $W$ dependence at $W\gtrsim 100$~GeV. The cross section for the case without proton shape fluctuations grows slightly more slowly as a function of energy compared to the case with proton substructure. This difference can be traced back to the fact  the parameter $\lqcd$ controlling the running coupling scale in coordinate space determined  in Ref.~\cite{Mantysaari:2022sux} 
is chosen to be smaller for the case using spherical nucleons, compared to the case where substructure fluctuations are included. This affects the evolution speed as a smaller $\lqcd$ leads to a smaller coupling constant. 

In order to quantify the magnitude of saturation effects in \jpsi photoproduction, we compute nuclear suppression factors separately for the coherent and incoherent channels. Following the definitions used in the recent experimental studies~\cite{ALICE:2023jgu, CMS:2023snh}, we define the suppression factor for the coherent production  as
\begin{equation}
    S_\mathrm{coh} = \sqrt{ \frac{\sigma^{\gamma A} }{ \sigma^\mathrm{IA} }}. \end{equation}
Here
    \begin{equation}
    \label{eq:impulseapprox}
        \sigma^\mathrm{IA} = \frac{\mathrm{d}\sigma^{\gamma p}}{\mathrm{d}t}(t=0) \int_{-t_\mathrm{min}} \!\!\!\! \dd{t} |F(t)|^2
    \end{equation}
is the corresponding cross section obtained from the impulse approximation (IA)~\cite{Chew:1952fca,Guzey:2013xba}, that is, the $\gamma+p$ result scaled to the $\gamma+\mathrm{Pb}$ case by only taking into account the nuclear form factor $F(t)$ and neglecting all other potential nuclear effects. 
In the LHC kinematics, we set $t_\mathrm{min}=0$.
We calculate the impulse approximation reference for the $\gamma+\mathrm{Pb}$ scattering exactly as the CMS collaboration: We use the approximate nuclear form factor $F(t)$ from Ref.~\cite{Klein:1999qj} with Woods-Saxon parameters $R_A=6.62$~fm and $a=0.535$~fm. When calculating $S_\mathrm{coh}$ for the gold nucleus to be compared with the upcoming STAR measurements, we use the same  Hartree–Fock–Skyrme nuclear density profile as STAR used e.g. in Ref.~\cite{Guzey:2013xba}, which corresponds to $\int_{-t_\mathrm{min}} \dd t |F(t)|^2=135.876$~GeV$^2$~\cite{kong_ia}.

When computing the nuclear modification factor for the incoherent cross section we follow the definition introduced by STAR~\cite{STAR:2023gpk,STAR:2023nos}:
\begin{equation}
    S_\mathrm{incoh} = \frac{ \sigma^{\gamma+ A \to \mathrm{J}/\psi + A^*}}{A (\sigma^{\gamma + p \to \mathrm{J}/\psi + p^*} + \sigma^{\gamma + p \to \mathrm{J}/\psi + p} ) }.
    \label{eq:sincoh}
\end{equation}
Note that unlike for the case of coherent production, the reference corresponds to the total diffractive (sum of coherent and incoherent) cross section in $\gamma+p$ scattering. Due to the different reference and the fact that there is a square root in the definition of $S_\mathrm{coh}$, one can not directly compare the two suppression factors, but we choose to adopt the same definitions as used in most currently available experimental studies. We also note that in general the incoherent cross section is expected to be more heavily suppressed: in the black disc limit where the fluctuations vanish, the incoherent $\gamma+A$ cross section vanishes, see Eq.~\eqref{eq:incohxs}, unlike the coherent production.

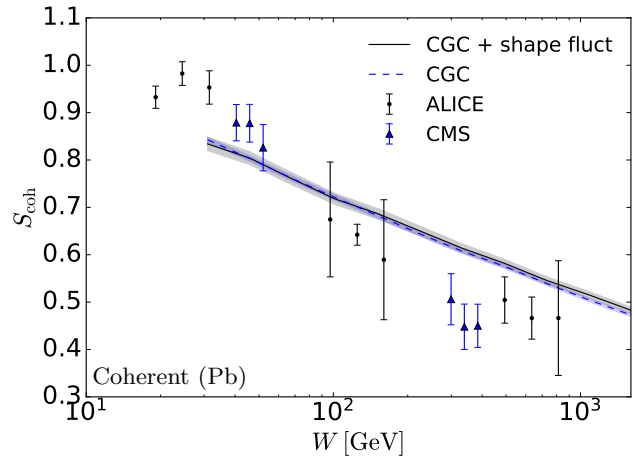
\begin{figure}

    \centering
    
    \includegraphics[width=\columnwidth]{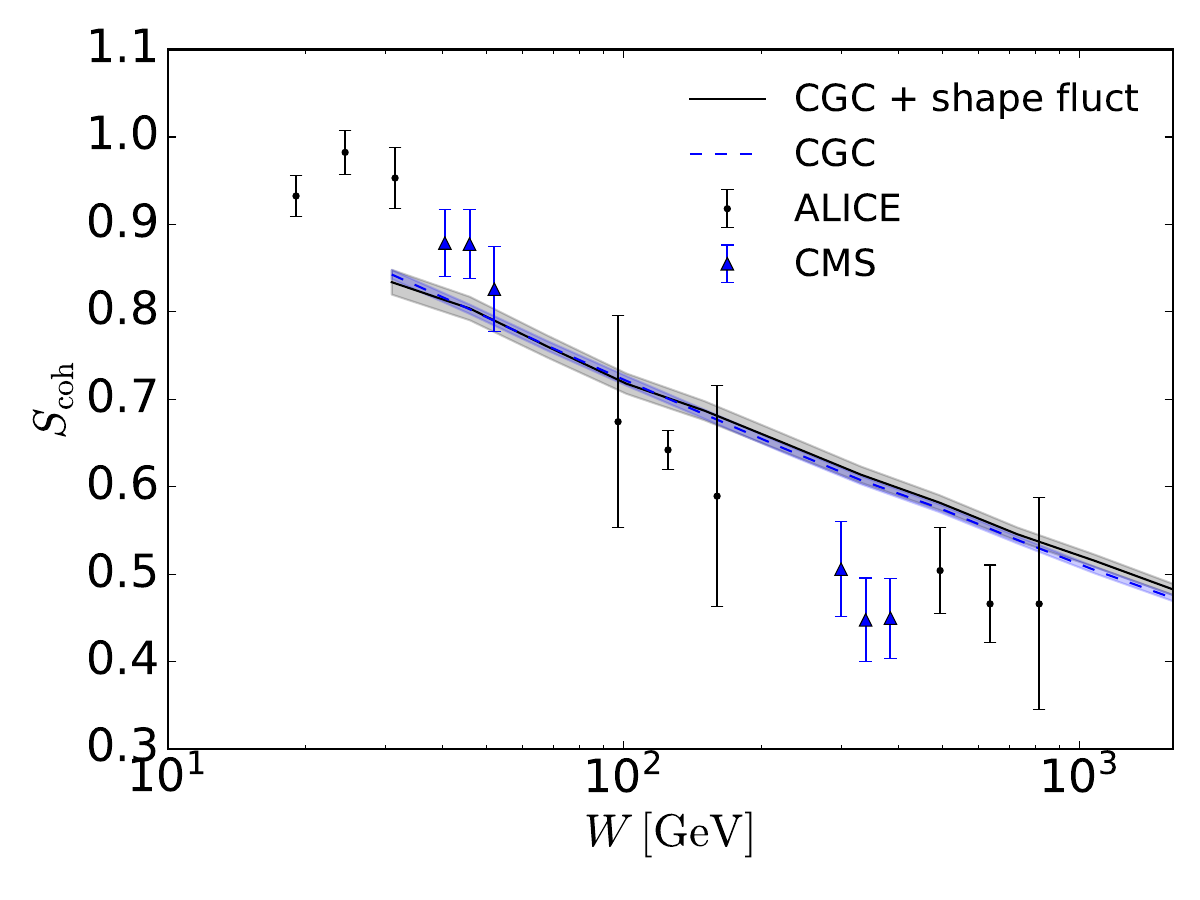}
    \begin{tikzpicture}[overlay,remember picture]
     \node at (-2.0cm,1.7cm){Coherent (Pb)};
     \end{tikzpicture}
            
            \caption{Suppression factor for coherent production compared to the ALICE~\cite{ALICE:2023jgu} and CMS data~\cite{CMS:2023snh}.
           }
            \label{fig:scoh}
\end{figure}

The obtained suppression factor for the coherent \jpsi photoproduction is shown in Fig.~\ref{fig:scoh}. Here we again show results calculated with and without nucleon substructure, and the same setup is used to compute both the $\gamma+\mathrm{Pb}$ process and the $\gamma+p$ reference. The results are compared to the ALICE~\cite{ALICE:2023jgu} and CMS~\cite{CMS:2023snh} data. 
We obtain slightly more suppression than the observed  $S_\mathrm{coh}\approx 0.9$ at the lowest center-of-mass energies $W\sim 45$~GeV, i.e., close to the initial condition of the JIMWLK evolution where the coherent photoproduction cross section was well reproduced as shown in Fig.~\ref{fig:photonuclearjpsi}. 
On the other hand, the suppression factor is overestimated for high center-of-mass energies.
Consequently, the $W$ dependence of the suppression factor is somewhat weaker in the employed CGC calculation compared to the LHC data.
This feature is reflected above in the fact that both the $\gamma+\mathrm{Pb}$ cross sections at high $W$ and the UPC cross section at $y=0$  (corresponding to $W=125$ GeV) are overestimated, but lower-energy data is better reproduced.
Note, however, that the impulse approximation baseline, Eq.~\eqref{eq:impulseapprox}, depends on the $\gamma+p\to\mathrm{J}/\psi+p$ cross section only at $t=0$ and not on the $t$-integrated cross section, which is experimentally better constrained, especially at high $W$. Consequently, the reference calculated from our setup is not precisely constrained by HERA data and there is some model uncertainty in the obtained suppression factors $S_\mathrm{coh}$. As  seen in Fig.~\ref{fig:coh_5020_nolhcb2021}, when nucleon substructure fluctuations are included a stronger nuclear suppression (smaller cross section) is obtained.
However, this effect is not visible in $S_\mathrm{coh}$ because the  $\gamma+p\to\mathrm{J}/\psi+p$ references differ at $t=0$ up to 10\% although the $t$-integrated cross sections are identical as constrained in Ref.~\cite{Mantysaari:2022sux}.

Comparisons to the STAR measurement of $S_\mathrm{coh}$ calculated using a gold target are shown in Table~\ref{table:suppression}. We calculate predictions at the initial condition of our JIMWLK evolution, $\xpom=0.01$, which is 
smaller than $\xpom=0.015$ probed in midrapidity measurements at STAR~\cite{STAR:2023gpk,STAR:2023nos}.
The JIMWLK evolution should not have a large effect in this small $\xpom$ range,
and we consider our predictions for $S_\mathrm{coh}$ to be relatively good approximations for STAR midrapidity kinematics. The STAR data is found to be compatible with our results.
Furthermore, by separating the high-$\xpom$ and low-$\xpom$ contributions to the UPC cross section, STAR may also be able to measure the cross section at $\xpom=0.01$.

\begin{table}[tb]
    \centering
    \begin{tabular}{c|ccc}
        Channel & STAR~\cite{STAR:2023nos,STAR:2023gpk}   & CGC + shape fluct  & CGC  \\ 
        \hline
                $S_\mathrm{coh}$ & $0.846 \pm 0.063$  & 0.89 & 0.90 \\
        $S_\mathrm{incoh}$ & $0.36^{+0.06}_{-0.07}$  & 0.58 & 0.32 \\
    \end{tabular}
    \caption{Nuclear modification factors for \jpsi photoproduction in $\gamma+\mathrm{Au}$ collisions. The CGC predictions are calculated at $\xpom=0.01$ and the STAR measurements are performed at $\xpom=0.015$.  The coherent suppression factors $S_\mathrm{coh}$ obtained with and without nucleon substructure fluctuations are compatible with each other within the numerical accuracy.}
    \label{table:suppression}
\end{table}

\begin{figure}
    \centering
        \includegraphics[width=\columnwidth]{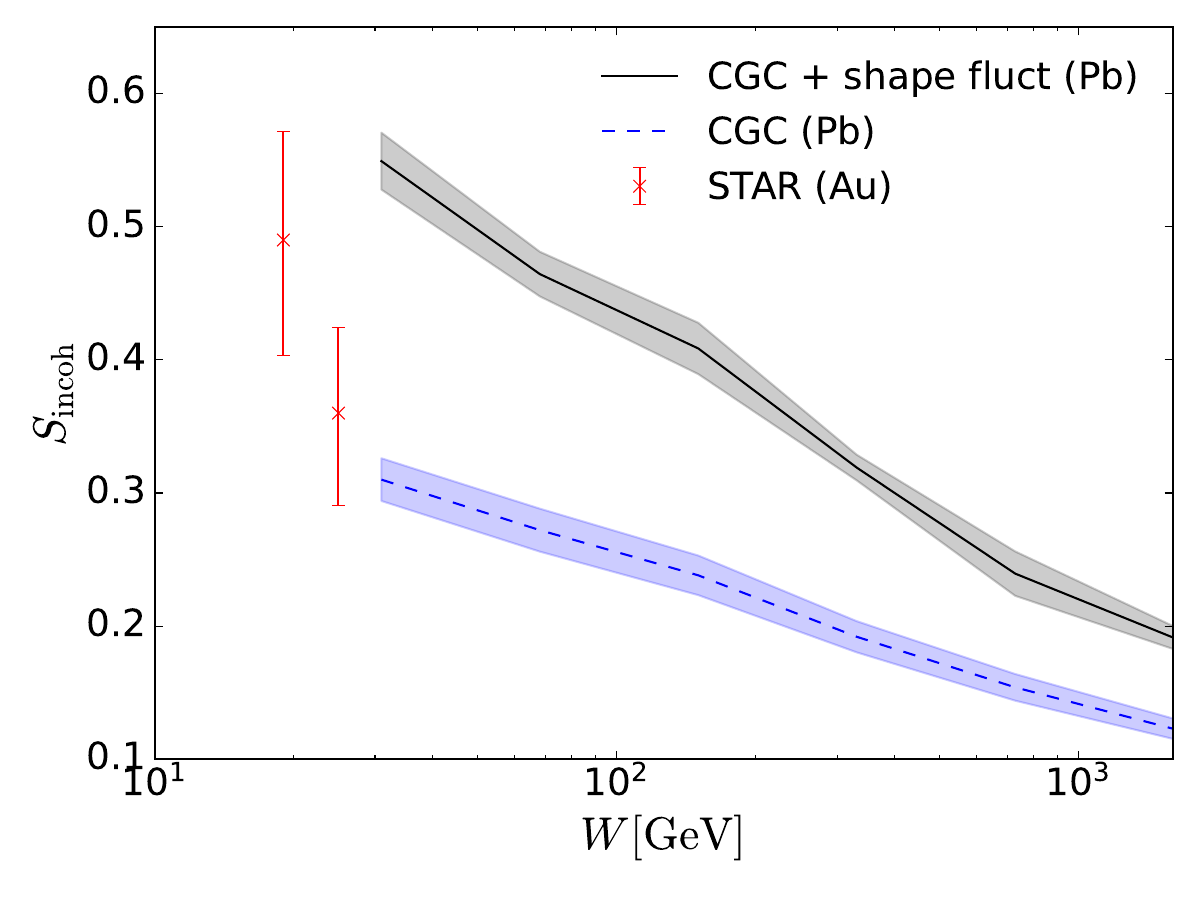}
         \begin{tikzpicture}[overlay,remember picture]
     \node at (-2.3cm,1.7cm){Incoherent};
     \end{tikzpicture}
        \caption{Energy dependence of $S_\mathrm{incoh}$ as defined in \eqref{eq:sincoh} for Pb nuclei calculated from the CGC setup and compared to STAR data~\cite{STAR:2023gpk,STAR:2023nos} for Au targets. The proton reference is always calculated with substructure fluctuations.}

    \label{fig:s_incoh}
\end{figure}
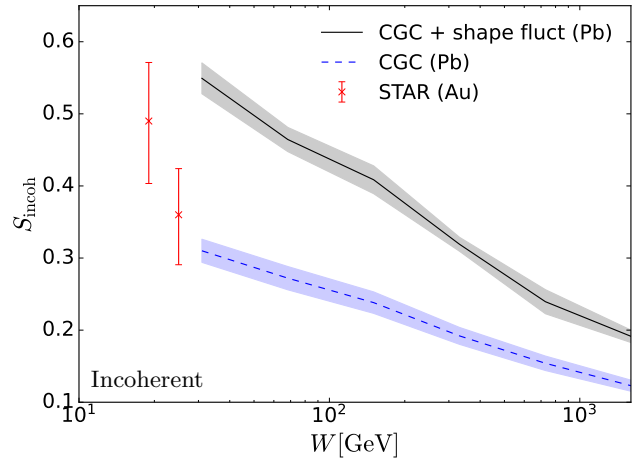

To complete the discussion of nuclear modification factors we present predictions for the suppression factor for the incoherent photonuclear \jpsi production defined in Eq.~\eqref{eq:sincoh}. The obtained suppression factors as a function of center-of-mass energy $W$ are shown in Fig.~\ref{fig:s_incoh} for $\gamma+\mathrm{Pb}$ collisions. For comparison, the STAR measurement for $\gamma+\mathrm{Au}$ collisions is shown~\cite{STAR:2023gpk,STAR:2023nos}. In order to use a $\gamma+p$ reference that is compatible with both the coherent and incoherent \jpsi production measurements at HERA, we include the proton shape fluctuations when calculating the denominator of Eq.~\eqref{eq:sincoh} independently of whether the nucleon shape fluctuations are included in the nucleus. 
Predictions for the $\gamma+\mathrm{Au}$ collisions in approximate STAR kinematics (calculated at $\xpom=0.01$, compared to STAR data at $\xpom=0.015$)  are shown in Table~\ref{table:suppression}.

When the nucleon shape fluctuations are included, the incoherent suppression factor is overestimated by $\sim 40\%$ at low $W$ in the STAR kinematics.
This is qualitatively consistent with the fact that the incoherent cross section in ultra peripheral collisions~\cite{ALICE:2013wjo} was found in Ref.~\cite{Mantysaari:2022sux} to be overestimated by $\sim 60\%$ at midrapidity LHC kinematics at $\sqrt{s}=2.76$~GeV. On the other hand, we note that in the $W$ range close to the STAR kinematics the obtained suppression factor for the coherent production is approximately compatible with the LHC data as shown in Fig.~\ref{fig:scoh}. If nucleon substructure fluctuations are not included for the nucleus, the suppression is overestimated.
This is because substructure fluctuations at short distance scales enhance the incoherent cross section significantly in the high-$|t|$ region~\cite{Mantysaari:2017dwh,Mantysaari:2022sux}. In our main setup with nucleon substructure included, we predict a faster $W$ dependence for $S_\mathrm{incoh}$ compared to the $S_\mathrm{coh}$, a genuine feature that can be tested with future LHC data. The STAR data hints at an even faster center-of-mass energy dependence than obtained in the setup with substructure fluctuations.
The strong suppression at high energies for the incoherent case is a result of JIMWLK evolution generating a smoother nucleus with less fluctuations and eventually approaching the black disc limit where the incoherent cross section vanishes and the coherent cross section dominates~\cite{Mantysaari:2018zdd}.

\subsection{Vector meson spectra}
\label{sec:incoh_spectra}

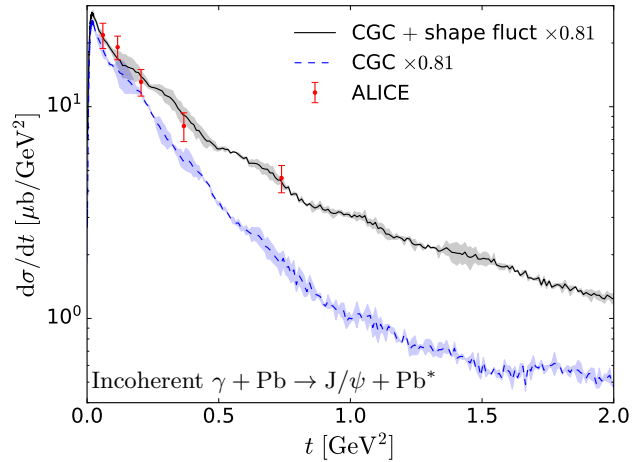
\begin{figure}
    \centering
    \includegraphics[width=\columnwidth]{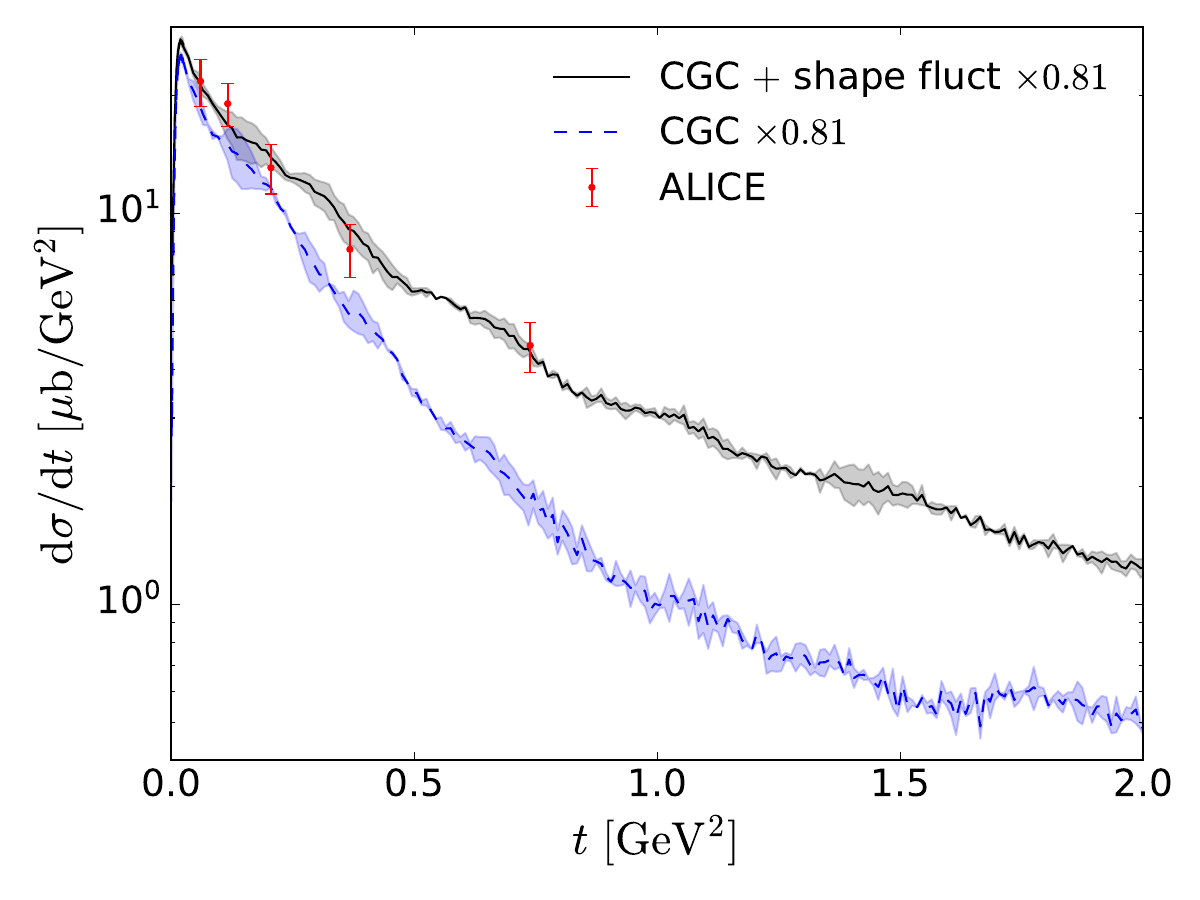}
    \begin{tikzpicture}[overlay,remember picture]
     \node at (-0.73cm,1.65cm){Incoherent $\gamma+\mathrm{Pb}\to\mathrm{J}/\psi+\mathrm{Pb}^*$};
     \end{tikzpicture}
    \caption{Incoherent \jpsi production in $\gamma+\mathrm{Pb}$ collisions at midrapidity compared to the ALICE data~\cite{ALICE:2023gcs}.}
    \label{fig:alice_incoh_t}
\end{figure}

To complete the discussion about the implications of new experimental UPC and $\gamma+A$ data that has become available since the publication of Ref.~\cite{Mantysaari:2022sux}, we calculate incoherent \jpsi production in  $\gamma$+Pb collisions as a function of squared momentum transfer. The results shown in Fig.~\ref{fig:alice_incoh_t} are compared with the ALICE data at $W=125$~GeV~\cite{ALICE:2023gcs} corresponding to midrapidity kinematics in UPCs at $\sqrt{s}=5020\,\mathrm{GeV}$. 
Based on Ref.~\cite{Mantysaari:2022sux} and the discussion above, we expect our incoherent cross section to overestimate the ALICE data. In order to better illustrate the shape of the $t$ distribution (which is sensitive to the substructure fluctuations) relative to the ALICE data, we show the results normalized by a factor $0.81$. This factor is determined by requiring that the ALICE data is optimally reproduced ($\chi^2$ is minimized) by the calculation that includes the nucleon substructure fluctuations. 
Note that the ($t$-integrated) coherent cross section shown in Fig.~\ref{fig:photonuclearjpsi} is overestimated by a slightly larger fraction at this kinematics: in order to reproduce the ALICE data point for the coherent cross section at $W=125$~GeV the same theory calculation (with substructure fluctuations) should be normalized by a factor $0.71$. This hints at a small tension between the coherent and incoherent data, but firm conclusions will require a precise measurement of the $t$-integrated incoherent cross section at this energy. We note that both the coherent and incoherent $t$-integrated UPC cross sections at $\sqrt{s}=2.76\,\mathrm{GeV}$ were consistently overestimated in Ref.~\cite{Mantysaari:2022sux}.

Relying on the description of the slope alone, we conclude that the ALICE data prefers the calculation with substructure fluctuations. Without such fluctuations, the calculated incoherent cross section decreases much faster in the $|t|\gtrsim 0.2$~GeV$^2$ region than the ALICE data. This is exactly the region where the $t$-slope is significantly modified and controlled by the size of the nucleon constituents that fluctuate~\cite{Lappi:2010dd,Demirci:2022wuy,Mantysaari:2017dwh,Toll:2021tcx,Mantysaari:2022sux}.
Similar conclusions supporting nucleon substructure fluctuations in nuclei based on comparisons to preliminary STAR data were reported in Ref.~\cite{Mantysaari:2022sux}.

\section{Conclusions}
\label{sec:conclusions}

We have calculated \jpsi photoproduction cross sections in photon-nucleus collisions and the corresponding nuclear modification factors within the Color Glass Condensate framework, where gluon saturation phenomena are naturally included. 
The experimentally measured coherent photoproduction cross section is well described in the range $30\,\mathrm{GeV}<W<50\,\mathrm{GeV}$ ($0.004<\xpom<0.01$). At high $W\gtrsim 100\,\mathrm{GeV}$, 
we reproduce the center of mass energy dependence of the data but overestimate the overall normalization. This suggests that the experimental data would prefer even stronger saturation effects at very high energies than what is obtained from our setup which is constrained by the $\gamma+p\to \mathrm{J}/\psi+p$ photoproduction data from HERA. This is also reflected by the fact that the nuclear suppression factor obtained for the coherent cross section is larger than what is seen in the ALICE and CMS data at high energies, and has a weaker dependence on the center-of-mass energy.

We have also calculated the nuclear suppression factor for the incoherent \jpsi photoproduction process. This is found to be highly sensitive to the nucleon substructure fluctuations in heavy nuclei. The only measurement available from STAR at low $W$ does not clearly prefer either a calculation with or without nucleon substructure fluctuations and as such leaves room for potential nuclear modification to the nucleon substructure within a heavy nucleus. 
The first data for the $t$-dependence of the incoherent \jpsi production from LHC is found to be compatible with no nuclear modification to the substructure fluctuations, although we again do not find large enough nuclear suppression. 
Future measurements for the energy dependence of the incoherent $\gamma+A\to\mathrm{J}/\psi+A^*$ cross section will make it possible to determine how the nucleon substructure fluctuations are modified by the saturation effects in heavy nuclei at high energies.

In the future, it will be important to consistently propagate the model uncertainties from fits to HERA $\gamma+p$ data (see e.g. Refs.~\cite{Mantysaari:2022ffw,Casuga:2023dcf}) to the calculations of high-energy $\gamma+\mathrm{Pb}$ scattering. 
Similarly, uncertainties originating from the non-perturbative vector meson wave function could be estimated by e.g. following Ref.~\cite{Lappi:2020ufv}.
This would allow one to determine if the strong nuclear suppression observed at the LHC can be described simultaneously with the $\gamma+p$ data where only weak saturation effects are expected~\cite{Armesto:2022mxy}. Furthermore, 
in order to achieve higher precision, all ingredients of the calculation should be advanced to next-to-leading order accuracy, see Refs.~\cite{Balitsky:2013fea,Kovner:2013ona,Beuf:2020dxl,Mantysaari:2021ryb,Mantysaari:2022kdm,Bendova:2020hkp,Iancu:2015vea,Ducloue:2019ezk,Beuf:2014uia,Hanninen:2022gje,Caucal:2023fsf} for related developments.
First estimates~\cite{Lappi:2021oag} indicated that the NLO corrections have only a small effect on the nuclear modification factor in exclusive vector meson production. However, recently it was found that nuclear modification in inclusive particle production in proton-nucleus collisions depends strongly on the initial condition chosen for the small-$x$ evolution equation~\cite{Mantysaari:2023vfh}, despite the fact that all these initial conditions have been constrained by the same proton structure function data~\cite{Beuf:2020dxl}. As such, the NLO effect on the nuclear modification factor studied in this work is currently an open question.
We emphasize that in the applied CGC setup there are no free parameters when moving from proton to nucleus and consequently the obtained nuclear suppression factor is a genuine prediction based on gluon saturation. This is in contrast to approaches based on collinear factorization where the nuclear modification to the (generalized) parton distribution function (PDF) is fit to data. Using nuclear PDFs determined from global analyses it is possible (within relatively large scale and PDF uncertainties) to get a good description of the nuclear suppression observed in UPCs~\cite{Eskola:2022vaf,Guzey:2020ntc}. Such global analyses including HERA and UPC data could also be performed within the CGC framework. If good agreement with data is achieved this method could provide powerful constraints on saturation effects in heavy nuclei.

\begin{acknowledgments}
We thank Z. Tu and Z. Ye for clarifying details of the STAR and CMS data and V. Guzey for useful discussions. 
H.M. is supported by the Research Council of Finland, the Centre of Excellence in Quark Matter, and projects 338263 and 346567, and under the European Union’s Horizon 2020 research and innovation programme by the European Research Council (ERC, grant agreement No. ERC-2018-ADG-835105 YoctoLHC) and by the STRONG-2020 project (grant agreement No. 824093) and wishes to thank the EIC Theory Institute at BNL for its hospitality during the completion of this work. B.P.S. is supported by the U.S. Department of Energy, Office of Science, Office of Nuclear Physics, under DOE Contract No.~DE-SC0012704  and within the framework of the Saturated Glue (SURGE) Topical Theory Collaboration. 
F.S. is supported in part by DOE under Contract No. DE-AC02-05CH11231, by NSF under Grant No. OAC-2004571 within the X-SCAPE Collaboration, and the INT’s U.S. DOE under Grant No. DE-FG02-00ER41132.
Computing resources from CSC – IT Center for Science in Espoo, Finland and the Finnish Grid and Cloud Infrastructure (persistent identifier \texttt{urn:nbn:fi:research-infras-2016072533}) were used in this work.
The content of this article does not reflect the official opinion of the European Union and responsibility for the information and views expressed therein lies entirely with the authors. 

\end{acknowledgments}
\bibliographystyle{JHEP-2modlong}
\bibliography{refs}

\end{document}